\documentclass[10pt,twoside]{article}
\usepackage{amsfonts}
\usepackage{fancyhdr}
\usepackage{titlesec}
\usepackage{cite}
\usepackage{ifthen}
\usepackage{amssymb}
\usepackage{pifont}
\usepackage{stmaryrd}
\usepackage{setspace}
\usepackage{indentfirst}
\usepackage{amsmath,amssymb,amscd,bbm,amsthm,mathrsfs,dsfont}
\input amssym.def

\usepackage{graphicx}

\titleformat{\section}{\centering\large\bfseries}{\S\arabic{section}}{1em}{}
\newboolean{first}
\setboolean{first}{true}

\begin{document}

\setlength\abovedisplayskip{2pt}
\setlength\abovedisplayshortskip{0pt}
\setlength\belowdisplayskip{2pt}
\setlength\belowdisplayshortskip{0pt}

\title{\bf \Large Numerical and Analytical Study of the Magnetic Field Distribution in a Three-Solenoid System \author{M. Behtouei $^{1}$, A. Bacci$^{1}$, M. Carillo$^{1}$, M. Comelli $^{2}$, L. Faillace$^{1}$,\\ M. Migliorati$^{3,4}$, L. Verra $^{1}$, B. Spataro$^{1}$, }\date{}} \maketitle
 %\footnote{Received: 2016-**-**.}%%%%%%% Please Input the time when the paper was submitted.
% \footnote{MR Subject Classification: .} %%%%%%% Please Input the according MR Classification Number.
%%%%%%%%%%%%%%%%% Please Input at least three keywords  %%%%%%%%%%%%%%

\noindent {
{$^{1}$  INFN, Laboratori Nazionali di Frascati, P.O. Box 13,
I-00044 Frascati, Italy}\\
{$^{2}$  {Istituto di Fisica Applicata "Nello Carrara" del Consiglio Nazionale delle Ricerche (CNR-IFAC), via Madonna del Piano 10, 50019, Sesto Fiorentino (FI)}\\
{$^{3}$ Dipartimento di Scienze di Base e Applicate per l'Ingegneria (SBAI), Sapienza University of Rome,\\Via Antonio Scarpa, 14, 00161 Rome, Italy}\\
{$^{4}$ INFN/Roma1, Istituto Nazionale di Fisica Nucleare, Piazzale Aldo Moro, 2, 00185, Rome, Italy}}

\indent  Email: Mostafa.Behtouei@lnf.infn.it

 %\footnote{Digital Object Identifier(DOI): 10.1007/s11766-017-****-*.}
% \footnote{Supported by the National Natural Science Foundation of China\,(no).}

\begin{abstract}
This paper investigates the magnetic fields produced by a three-coil system, focusing on how different mesh resolutions affect the accuracy of the results. Using both the Poisson solver, as well as a numerical approach based on the solution of fractional integrals, the study examines coils with dimensions of 80 mm by 160 mm and a radius of 15.5 mm, each carrying a current of 200 A. The research explores how varying mesh step sizes within the coil regions influences the simulations accuracy and convergence. The results are analyzed along a line parallel to the central axis at a distance equal to half of the solenoid's radius.. The findings emphasize the consistency of the numerical results, the compatibility of the solvers, and offer insights into further optimization strategies for more efficient simulations.

\end{abstract}

\section{Introduction}
Accurate modeling of magnetic fields in systems with multiple coils is essential for the design and optimization of advanced electromagnetic devices. The ability to precisely determine magnetic fields in solenoids and coils plays a central role in many areas of applied physics and engineering. These field components are fundamental in the development of particle accelerators such as high-brilliance photoinjectors and high-gradient accelerating structures as well as in magnetic correctors and steering systems used to align particle beams. A key challenge in these applications is the control and stabilization of beam trajectories, particularly in the presence of space charge effects \cite{ferrario2007direct,migliorati2009transport,stratakis2010effects,nezhevenko200134,brady2013encyclopedia,behtouei2020initial,behtouei2021ka}. In this work, we examine a system composed of three coils, exploring how mesh resolution impacts numerical accuracy. The study analyzes the magnetic field behavior in off-axis positions using the Poisson solver \cite{mckenney1995fast}. Our goal is to validate the numerical models and quantify potential errors arising from different mesh configurations.

The theoretical basis for calculating magnetic fields generated by current-carrying conductors is well established. At its core lies the Biot-Savart law, a direct consequence of Maxwell's equations, which provides a powerful tool for evaluating magnetic fields from a given current distribution \cite{jackson1999classical}. This law has enabled analytical solutions for idealized cases such as thin wires, loops, and infinite solenoids. However, when dealing with more realistic geometries like solenoids of finite length or conductors with distributed surface currents, the problem becomes considerably more complex and often requires numerical approaches.

Over the years, several researchers have proposed methods to overcome these challenges. For example, Bassetti et al. \cite{bassetti1996analytical} developed an analytical model for the near-axis magnetic potential of multipole configurations, which relied heavily on numerical techniques to handle higher-order terms when moving away from the axis. Similarly, Caciagli et al. \cite{caciagli2018exact} derived analytical expressions for the magnetic field of finite-length cylinders with uniform transverse magnetization using complete elliptic integrals. Despite the progress, the presence of fractional integrals of order 3/2 and 1/2 in these formulations introduced mathematical complications, such as branch lines, which required special attention in the computations.

In our previous work \cite{behtouei2020novel,behtouei2022novel}, we tackled these difficulties by analyzing solenoids composed of discrete coils and introduced an innovative method to address the issue of branch lines in fractional integrals. We later extended this approach to solenoids formed by continuous conductive sheets, using surface current density distributions to derive the corresponding magnetic fields \cite{behtouei2020novel}.

We begin by analyzing the magnetic field produced by a single thin coil of radius R, evaluating its intensity both along the axis and at various off-axis locations. We then apply the superposition principle to construct the field of a complete solenoid. Through a detailed comparison between analytical and numerical solutions, we aim to provide a deeper understanding of the field characteristics across different configurations.

By integrating theory and simulations, this study offers a comprehensive framework for accurately determining magnetic fields in solenoids. The results have significant implications for accelerator physics, beam dynamics, and the design of magnetic devices in cutting-edge scientific and technological applications.

\section{Analytical method}

The determination of magnetic fields in solenoids and coils can be addressed through both analytical and numerical approaches. In this section, we present the key analytical expressions derived in previous works, focusing on the magnetic field components for different configurations. These results serve as a foundation for the subsequent numerical validation.

\subsection*{Magnetic field for a single thin coil}

For a single thin circular coil of radius \( R \), carrying a steady current \( I \), the magnetic field components can be derived using the Biot-Savart law. The resulting expressions for the radial (\( B_r \)) and axial (\( B_z \)) components are given by \cite{behtouei2020novel}:

\begin{equation}\label{55}
\mathrm{B}_r = \frac{\mu_0 I M_z}{2 R} \left(\frac{\xi}{2\eta(1+\xi)}\right)^{3/2}
\left[\ _2F_1\left(\frac{3}{2}, \frac{3}{2}; 2; \frac{2\xi}{1+\xi}\right) - \ _2F_1\left(\frac{1}{2}, \frac{3}{2}; 1; \frac{2\xi}{1+\xi}\right)\right],
\end{equation}

\begin{equation}
\mathrm{B}_z = \frac{\mu_0 I}{2 R} \left(\frac{\xi}{2\eta(1+\xi)}\right)^{3/2}
\left[(1+\eta) \ _2F_1\left(\frac{1}{2}, \frac{3}{2}; 1; \frac{2\xi}{1+\xi}\right) 
- \eta \ _2F_1\left(\frac{3}{2}, \frac{3}{2}; 2; \frac{2\xi}{1+\xi}\right)\right],
\end{equation}

where \(\mu_0\) represents the permeability of free space, \(I\) is the current passing through the coil, and \(R\) is the radius of the coil. The normalized radial and axial distances are given by \(\eta = \frac{r}{R}\) and \(M_z = \frac{z}{R}\), respectively, and \( _2F_1(a,b;c;z) \) denotes the Gaussian hypergeometric function. The dimensionless parameter \(\xi\), which captures the geometry of the field configuration, is defined as

\begin{equation}
\xi(r, R, z) = \frac{2rR}{r^2 + R^2 + (z - z_0)^2},
\end{equation}

\subsection*{Magnetic field for finite-length solenoids}

For a solenoid with a finite length, the magnetic field becomes more complex due to the extended current distribution along the axial direction. Assuming a uniform surface current density \( K \), the radial and axial components of the magnetic field can be expressed as \cite{behtouei2022novel}:

\begin{multline}\label{2.14}
\mathrm{B}_r = \frac{\mu_0 K}{2} \int_{\sqrt{(1+\eta)^2 + (M_z-M_{z_\ell})^2}}^{\sqrt{(1+\eta)^2 + (M_z+M_{z_\ell})^2}} \frac{1}{\beta^2} 
\left[\ _2F_1\left(\frac{1}{2}, \frac{3}{2}; 1; \frac{4\eta}{\beta^2}\right) \right. \\
\left. - \ _2F_1\left(\frac{3}{2}, \frac{3}{2}; 2; \frac{4\eta}{\beta^2}\right)\right] \, d\beta,
\end{multline}

\begin{multline}\label{2.15}
\mathrm{B}_z = \frac{-\mu_0 K}{2} \int_{\sqrt{(1+\eta)^2 + (M_z-M_{z_\ell})^2}}^{\sqrt{(1+\eta)^2 + (M_z+M_{z_\ell})^2}} 
\frac{1}{\beta^2 (\beta^2 - (1+\eta)^2)^{1/2}} \\
\left[\ _2F_1\left(\frac{1}{2}, \frac{3}{2}; 1; \frac{4\eta}{\beta^2}\right) 
+ \eta \left(\ _2F_1\left(\frac{1}{2}, \frac{3}{2}; 1; \frac{4\eta}{\beta^2}\right) 
- \ _2F_1\left(\frac{3}{2}, \frac{3}{2}; 2; \frac{4\eta}{\beta^2}\right)\right)\right] \, d\beta.
\end{multline}

Here, \( M_{z_\ell} = \frac{L}{2R} \) (with \( L \) being the solenoid length), and \( \beta \) is the integration variable corresponding to the radial distance from the axis. These integrals represent the contributions from all the current loops composing the solenoid.

These analytical expressions provide the benchmark against which numerical simulations will be compared in the following sections.

\section{Comparison between numerical simulations and analytical results}

The layout and design of the three coils used in this study are illustrated in Figure 1, showing the geometric configuration and structural considerations \cite{vannozzi}. Each coil is designed with a rectangular cross-section measuring 80 mm by 160 mm. The coils are constructed with 10 longitudinal layers and 20 radial layers, each made from square wire segments with dimensions of 7.5 mm by 7.5 mm. To ensure efficient cooling during high-current operations, each wire segment is equipped with a central hole of 5 mm diameter, facilitating coolant circulation through the coil. Moreover, a 0.5 mm gap is maintained between adjacent wire segments to provide insulation, helping to minimize thermal stress and maintain the structural integrity of the coils. 
\begin{figure}[h!]
 \begin{center}
\includegraphics[width=0.7\linewidth]{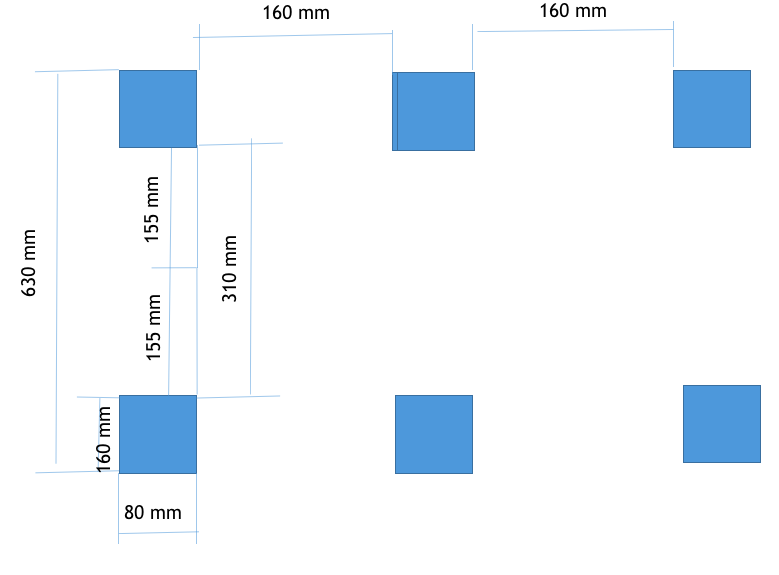}
\caption{Schematic layout of the three-spool solenoid configuration. Each spool has a width of 80 mm and a height of 160 mm, with a vertical spacing of 155 mm between adjacent spools. The total height of the solenoid system is 630 mm. The horizontal distance between neighboring solenoids is 160 mm, and the central solenoid is positioned symmetrically between the two outer ones.}
\end{center}
\end{figure}

The coils are arranged symmetrically in a rectangular layout, with a vertical separation of 155 mm and horizontal spacing of 160 mm. This results in a total system height of 630 mm and a width of 310 mm. This configuration not only ensures a uniform magnetic field distribution but also integrates essential thermal management features to support the coils' operation at a total current of 40 kA each. The design effectively combines efficiency and reliability, making it suitable for high-power applications.

\begin{figure}
 \begin{center}
\includegraphics[width=0.7\linewidth]{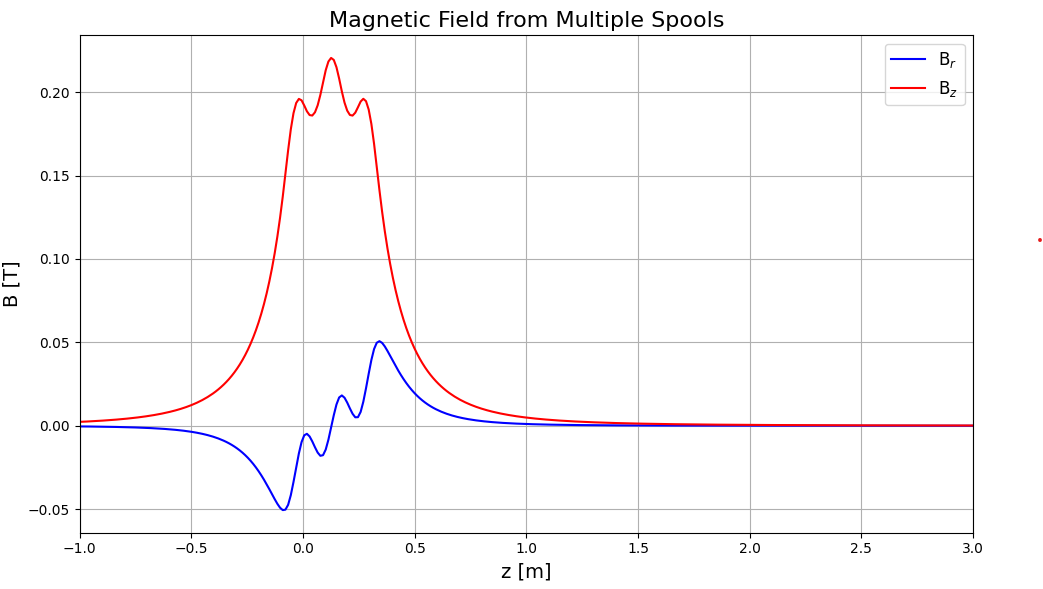} a)
\includegraphics[width=0.7\linewidth]{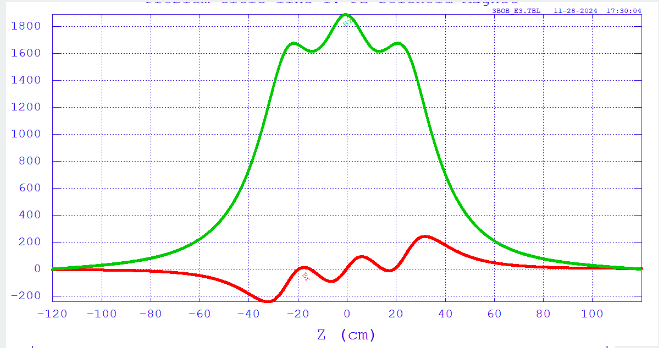} b)
\includegraphics[width=0.7\linewidth]{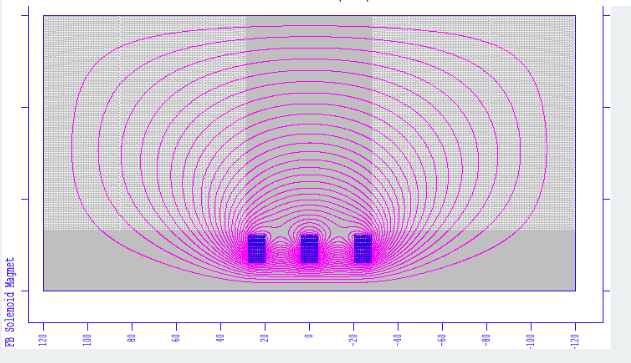} c)
\caption{ Analytical and numerical results of the axial and radial magnetic field generated by three solenoid spools at a off-axis distance of 0.5 R.  a) Analytical magnetic field results.  b) Simulations results obtained using Poisson software.  c) Magnetic field lines of the solenoid, simulated using Poisson software. }
\end{center}
\end{figure}
Magnetic field simulations for this coil system was carried out using the Poisson solver, both of which are highly effective for modeling electromagnetic systems with high precision. The simulations were performed using three different mesh resolutions 200 $\mu$m, 400 $\mu$m, and 590 $\mu$m to ensure that field dynamics were captured accurately while balancing computational resource usage. Within the winding regions of the coils, a uniform and fine mesh (\(dx = dy\)) was used to achieve high resolution where magnetic field gradients are expected to be most significant. In contrast, a coarser mesh was applied outside the coils, where magnetic field variations are less critical, allowing for reduced computational overhead without sacrificing overall simulation accuracy.

This combination of fine and coarse meshing strategies provides a reliable method for balancing computational efficiency with the need for detailed electromagnetic analysis. The results from these simulations were compared against theoretical predictions to assess the accuracy of the field distribution.

Magnetic field simulations are essential for the design and analysis of electromagnets and solenoids, particularly when aiming to balance numerical accuracy with computational efficiency. This section examines the trade-off between these two factors, specifically focusing on the accuracy of magnetic field calculations obtained through the Poisson solver.

Figure 2 presents the analytical and numerical results of the magnetic field generated by a solenoid composed of three spools. Figure 2a shows the magnetic field components, both axial and radial, at an off-axis distance of 0.5 R (R = 155 mm), obtained through analytical methods.  Figure 2b presents the corresponding magnetic field components at the same off-axis distance derived from  Poisson software. Figure 2c depicts the magnetic field lines of the solenoid, as simulated by Poisson software, offering a visual representation of the field's direction and intensity across the solenoid's structure. 

\begin{figure}
 \begin{center}
\includegraphics[width=0.7\linewidth]{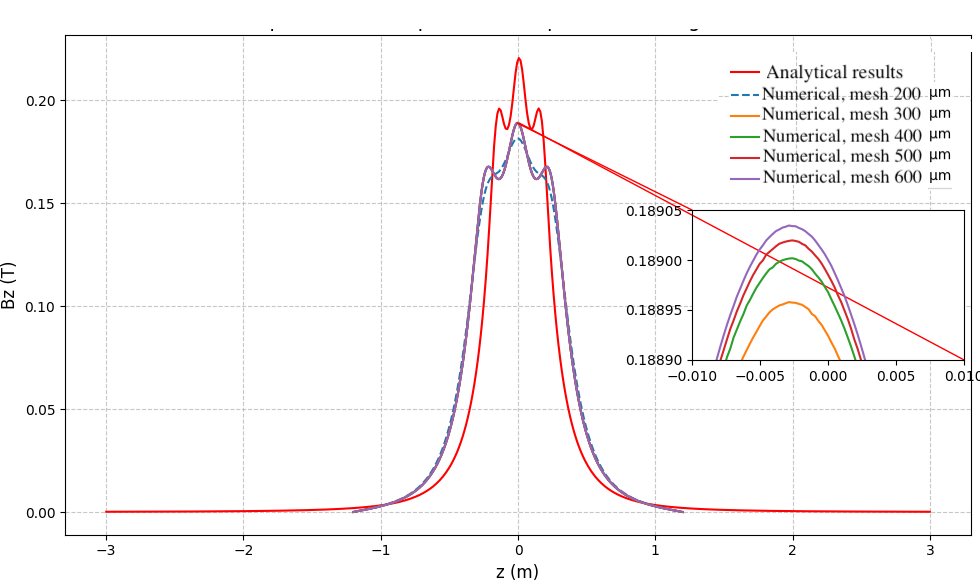} a)
\includegraphics[width=0.7\linewidth]{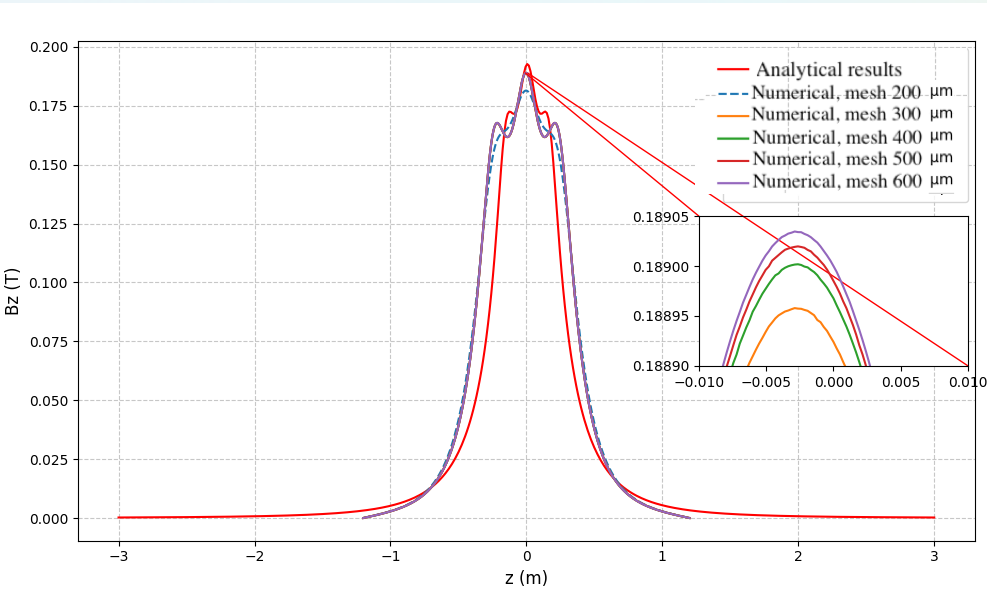} b)
\includegraphics[width=0.7\linewidth]{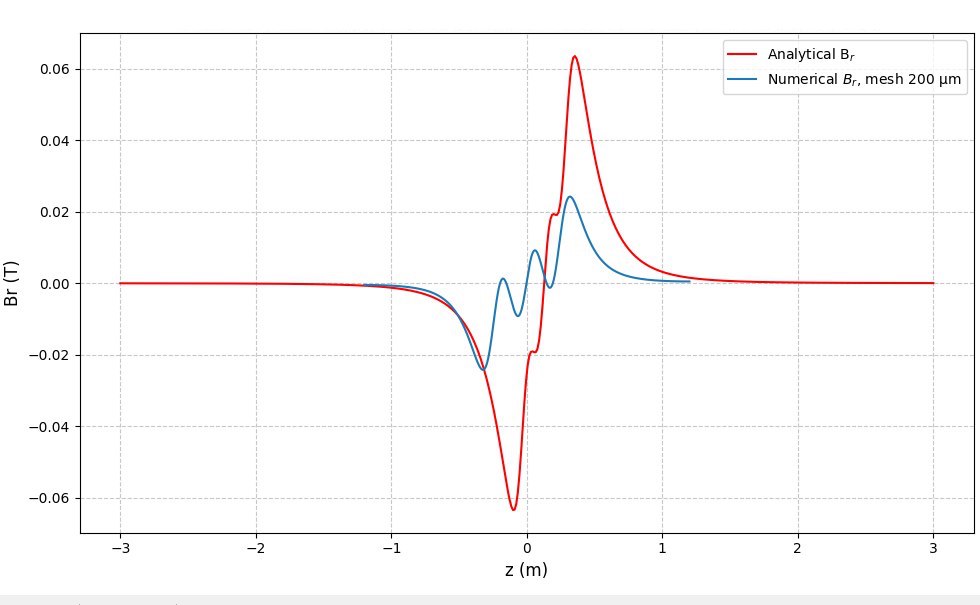} c)
\caption{Comparison of analytical and numerical magnetic field results for the three solenoid spools. (a) Axial field with coil radius 155 mm. (b) Axial field with average radius 235 mm. (c) Radial field with average radius 235 mm and 200 $\mu$m mesh size.}
\end{center}
\end{figure}

Figure 3 shows a comparison between analytical and numerical results. Figure 3a presents the axial magnetic field using a coil radius of 155 mm, with different mesh sizes in simulations compared to the analytical solution. The small plot in the corner helps to better see the differences near the center of the structure. Figure 3b shows a similar comparison for the axial magnetic field, but using an average coil radius of 235 mm, which represents the average radius of the solenoid's coils \cite{Lyle}. The results still show good agreement between the two methods. Figure 3c shows the radial magnetic field for the average coil radius of 235 mm, comparing numerical and analytical results using a 200 $\mu$m mesh size. The graph shows how the radial field changes near the edges of the solenoid. While the axial magnetic field obtained from numerical simulations closely matches the analytical solution, the radial component exhibits a significant discrepancy by a factor of approximately 2, when comparing the two methods. This difference is likely due to the increased sensitivity of the radial field to geometric approximations and edge effects inherent in the numerical model, as discussed by Fleurot et al. \cite{fleurot2021analytical}

The mesh resolution plays a crucial role in determining the accuracy of the magnetic field simulations. A finer mesh improves the precision of the results but increases the computational cost, while a coarser mesh reduces the computational time but may introduce errors. Table \ref{tab:mesh_resolution} summarizes the peak magnetic field values (\(B_z\)) and corresponding computation times for the various mesh resolutions tested in this study.

\begin{table}[h!]
\centering
\begin{tabular}{|c|c|c|}
\hline
\textbf{Mesh Size (\(\mu m\))} & \textbf{Peak \(B_z\) (G)} & \textbf{Computation Time (h:min)} \\
\hline
200 & 1889.6 & 4:00 \\
300 & 1890.0 & 2:30 \\
400 & 1890.2 & 1:30 \\
590 & 1890.4 & 0:30 \\
\hline
\end{tabular}
\caption{Peak magnetic field values and computation times for different mesh resolutions.}
\label{tab:mesh_resolution}
\end{table}

As shown in Table \ref{tab:mesh_resolution}, the differences in peak \(B_z\) values across the mesh resolutions are minimal, with a maximum deviation of less than \(0.05\%\). This demonstrates the numerical stability of the solver, even with varying mesh sizes. While the computational time decreases significantly with coarser meshes, the \(590 \, \mu m\) mesh offers the most efficient balance, providing results comparable to those from finer meshes but at a fraction of the computational cost.

To further assess numerical accuracy, the relationship between mesh resolution and the magnetic field estimations (\(B_z\)) was analyzed. Figure~4b presents a second-degree polynomial fit of the form:

\[
Y = M_0 + M_1 x + M_2 x^2,
\]

with coefficients \(M_0 = 1888.6\), \(M_1 = 0.0063478\), and \(M_2 = -5.4859 \times 10^{-6}\). The high coefficient of determination (\(R = 0.99793\)) reflects an excellent fit to the data, with the maximum error between the actual data and the polynomial estimation being less than \(0.003\%\). This demonstrates the robustness of the polynomial model and validates its use for approximating field behavior as a function of mesh size.

Figure 4a shows the same data fitted using a third-degree polynomial. This model captures the slight nonlinear behavior more effectively as the mesh becomes finer, resulting in an even better fit with a correlation coefficient of \(R = 1\). The third-degree polynomial, shown in Figure4a, thus provides a more precise approximation, especially when the changes in the magnetic field are not perfectly smooth. Together, the two fits confirm the consistency and accuracy of the numerical approach across different polynomial models and mesh resolutions.

\begin{figure}[h!]
\centering
\includegraphics[width=0.4\textwidth]{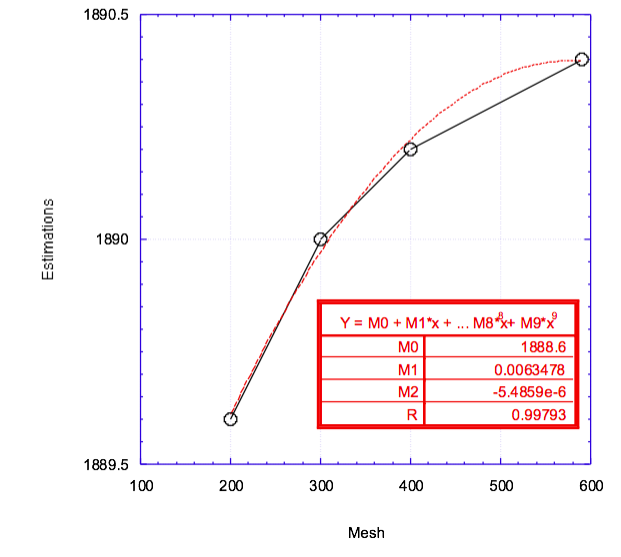} a)
\includegraphics[width=0.4\textwidth]{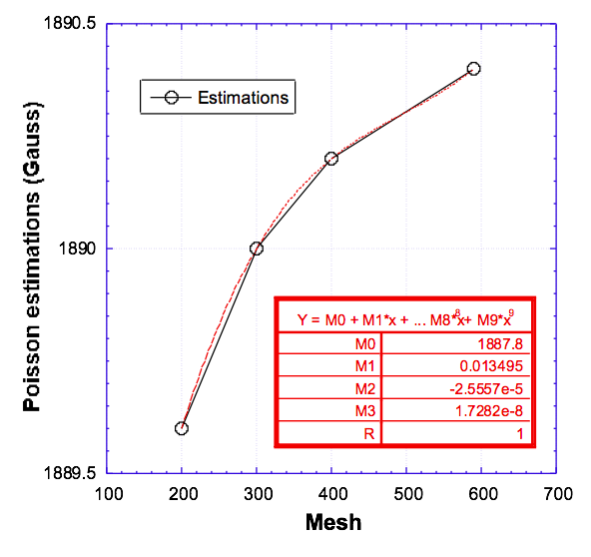} b)
\caption{a) Poisson field estimations (\(B_z\)) as a function of mesh size, with a polynomial fit for error analysis. b) Poisson field estimations (\(B_z\)) as a function of mesh size, analyzed with a second-degree polynomial fit for error estimation. The approximation error is shown to be less than 0.003\%.}
\end{figure}

The numerical error is defined as the difference between the numerical solution \(\mathbf{B}_h\) and the exact solution \(\mathbf{B}_{\text{exact}}\), with the error function given by:

\[
\text{Error}_{L_2}(h) = \sqrt{\int_\Omega \|\mathbf{B}_h - \mathbf{B}_{\text{exact}}\|^2 \, d\Omega},
\]

where \(\Omega\) is the computational domain and \(h\) is the mesh size. The error inside the solenoid is given by \cite{bathe2006finite}:

\[
\text{Error}_{\text{inside}} = C h^p \sqrt{\pi R^2},
\]

where \(R\) is the solenoid radius, and the error outside the solenoid is negligible because the exact magnetic field there is nearly zero, making the numerical error minimal. Thus, its contribution to the total $L_2$ norm error is insignificant compared to the inside. The total error is dominated by the contribution inside the solenoid, and it scales as \(\text{Error}_{L_2}(h) \propto h^p\), indicating that smaller mesh sizes or higher-order numerical methods are needed for higher accuracy \cite{reddy2019introduction,bathe2006finite}.

The agreement between the Poisson and analytical method further validates the reliability of the simulations. Future work could investigate advanced solver algorithms or adaptive mesh refinement techniques to further enhance accuracy while reducing computational costs.

\section{Conclusion}
This study demonstrates the high accuracy achieved in magnetic field simulations for a three-coil system, even when different mesh resolutions are applied. The results show that the numerical error is exceptionally low across various mesh sizes, with the peak axial magnetic field, \(B_z\), varying by less than 0.05\%. This suggests that the simulations are highly reliable, even when using coarser meshes. Furthermore, the agreement between the Poisson solver and analytical method was robust, further confirming the validity of the simulation approach, especially when using simpler, less computationally expensive mesh configurations. 

The findings also provide valuable guidance for selecting the appropriate mesh resolution, ensuring a good balance between computational efficiency and the level of precision required for accurate field analysis.

\end{document}